\documentclass[runningheads]{llncs}
\usepackage[T1]{fontenc}
\usepackage{makecell}
\usepackage{multirow}
\usepackage{amsmath}
\usepackage{amssymb}
\usepackage{graphicx}
\usepackage{subfig}
\usepackage{color}
\title{HarDNet-DFUS: An Enhanced Harmonically-Connected Network for
Diabetic Foot Ulcer Image Segmentation and
Colonoscopy Polyp Segmentation}
\begin{document}

\author{Ting-Yu Liao $^*$ \and Ching-Hui Yang $^*$ \and Yu-Wen Lo $^*$ \and \\ Kuan-Ying Lai \and Po-Huai Shen \and Youn-Long Lin}
\authorrunning{ }
\titlerunning{ }

\institute{Department of Computer Science, National Tsing Hua University, Hsinchu, TAIWAN
\email{\{wendy107062324, hui09080729, wagw1014, kytimmylai\}@gapp.nthu.edu.tw,
ty2134029@gmail.com,
ylin@cs.nthu.edu.tw}}

\maketitle              % typeset the header of the contribution

\def\thefootnote{*}\footnotetext{These authors contributed equally to this work}

\begin{abstract}
    We present a neural network architecture for
    medical image segmentation of diabetic foot ulcers and 
    colonoscopy polyps.
    Diabetic foot ulcers are caused by neuropathic and vascular complications of diabetes mellitus.
    % It remains a major risk factor for foot osteomyelitis and lower extremity amputations. 
    In order to provide a proper diagnosis and treatment,
    wound care professionals need to extract accurate morphological features
    from the foot wounds.
    % However, the process of visual and manual inspection is very time consuming and error-prone. 
    Using computer-aided systems is a promising approach
    to extract related morphological features and segment the lesions.
    We propose a convolution neural network called HarDNet-DFUS by
    enhancing the backbone and replacing the decoder of
    HarDNet-MSEG, which was SOTA for colonoscopy polyp segmentation in 2021.
    For the MICCAI 2022 Diabetic Foot Ulcer Segmentation Challenge (DFUC2022),
    we train HarDNet-DFUS using the DFUC2022 dataset and
    increase its robustness
    by means of five-fold cross validation, Test Time Augmentation, etc.
    In the validation phase of DFUC2022,
    HarDNet-DFUS achieved 0.7063 mean dice and
    was ranked third among all participants.
    In the final testing phase of DFUC2022,
    it achieved 0.7287 mean dice and
    was the first place winner.
    HarDNet-DFUS also deliver excellent performance for
    the colonoscopy polyp segmentation task.
    It achieves 0.924 mean Dice on the famous Kvasir dataset,
    an improvement of 1.2\% over the original HarDNet-MSEG.
    The codes are available on
    https://github.com/kytimmylai/DFUC2022
    (for Diabetic Foot Ulcers Segmentation) and https://github.com/YuWenLo/HarDNet-DFUS 
    (for Colonoscopy Polyp Segmentation).
    

\end{abstract}
\section{Introduction}
  % illustrate the impact
  Diabetes is a global epidemic, and it is estimated that by the end of 2045, approximately 600 million people will have diabetes.
  Diabetic Foot Ulcers (DFU) is one of the complications of diabetes,
  often leading to more serious conditions such as infection and ischemic,
  which can significantly prolong treatment and
  often lead to amputation and, in more severe cases, death.
  % why we need the tech
  In current practice, medical professionals primarily
  use manual measurement tools to visually examine and
  evaluate patients with DFU to determine its severity.
  However, this is not only time consuming but also
  challenging for podiatrists.
  Therefore, it becomes important to
  fast and accurately determine the exact region of the ulcer.
  
  % development of the tech
  In recent years, based on the rapid development of
  convolutional neural networks,
  many deep learning techniques have been applied
  in the field of medical imaging. 
  % related work
  For this task, U-Net \cite{U-Net} employs
  an encoder-decoder architecture
  that has achieved breakthrough performance and
  stimulated many improvements,
  such as ResUNet++ \cite{ResUNet++}, DoubleU-Net \cite{DoubleU-Net}, UNet++ \cite{UNet++}, etc. 
  % what's the goal?
  However, the overly complex network architecture,
  low accuracy of small target segmentation, and slow segmentation speed have limited the practical deployment of U-Net variants in the clinical field.
  
  % what's new
  Therefore, based on the previously state-of-the-art
  HarDNet-MSEG \cite{HarDNet-MSEG} for colonoscopy polyp segmentation,
  we enhance its backbone incorporating the concept of CSPNet \cite{CSPNet} and ShuffleNetV2 \cite{ShuffleNetV2}, and
  employ a new decoder introduced in
  the Lawin Transformer\cite{lawin}.
  We called the resultant network HarDNet-DFUS.
  It improves the capability of detecting ulcer regions and
  can deliver better accuracy compared to
  the original HarDNet-MSEG.  
  
  % summary
  The contributions of this study can be summarized as follows: 
  First, we have improved the HarDNet-MSEG model to
  achieve better performance in ulcer region segmentation.
  Second, we have enhanced the HarDNet \cite{HarDNet} backbone
  to achieve higher accuracy while keeping a similar speed.
  Third, we have evaluated the proposed method using
  the single-class segmentation tasks of the DFUC 2022 Challenge.
\section{Method}

\begin{figure}[!ht]
    \centering
     \subfloat[]{\includegraphics[width=8cm]{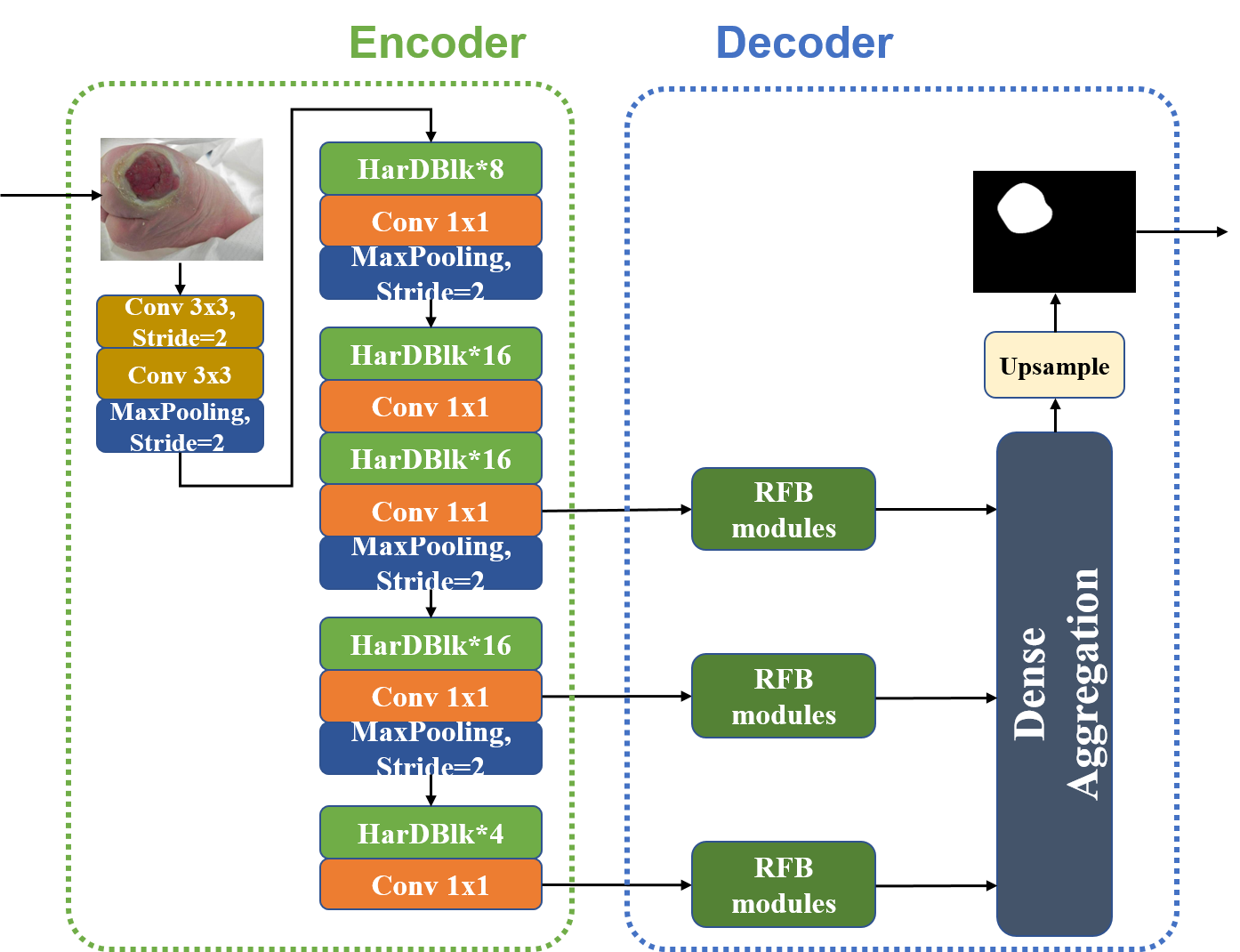}\label{fig:sub_ref_3}}
    \hspace{1pt}
    \centering
     \subfloat[]{\includegraphics[width=12cm]{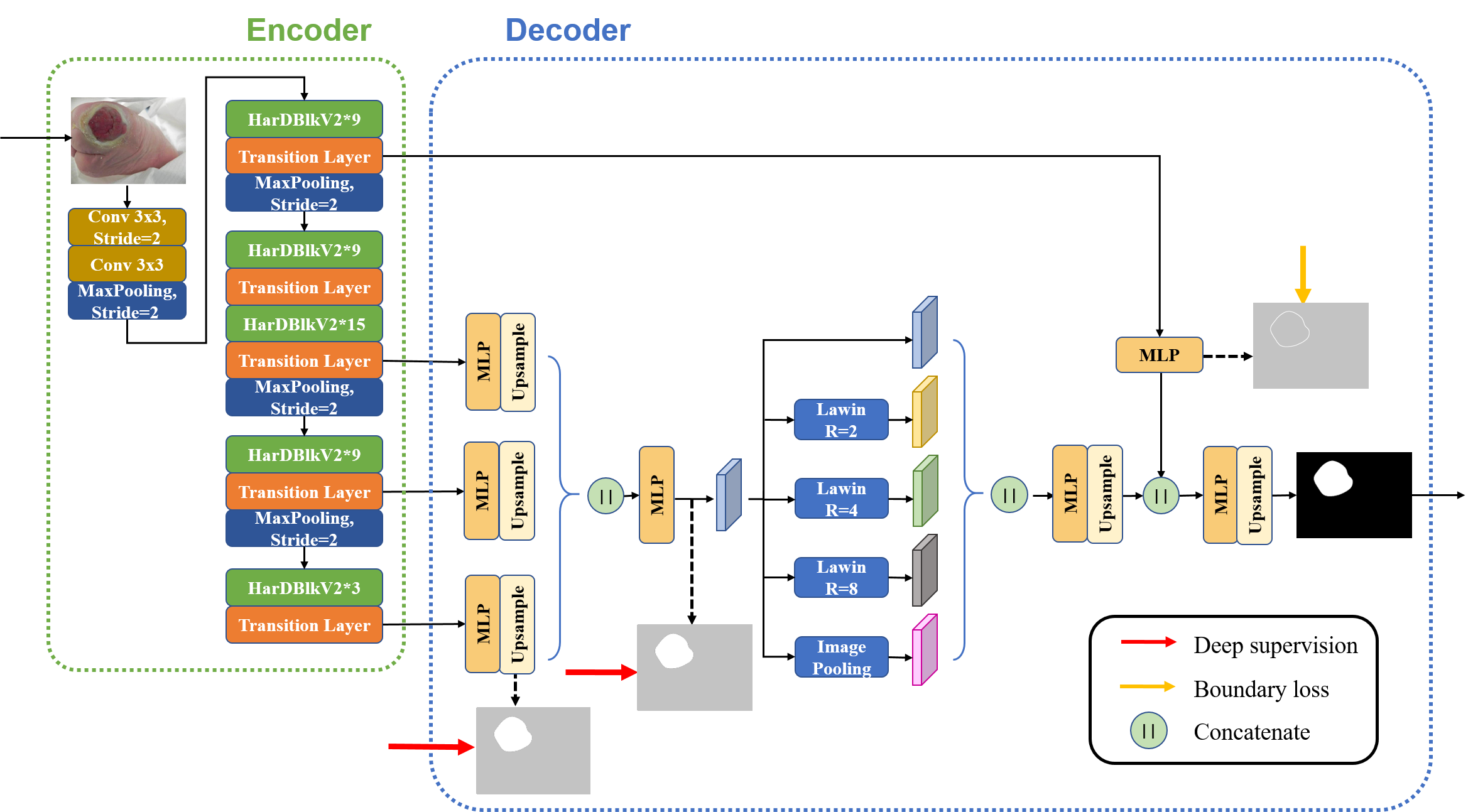}\label{fig:sub_ref_4}}
    %% replacing -> modifying
    \caption{(a) Original HarDNet-MSEG model, (b) Enhanced model by replacing HarDBlock with HarDBlockV2 and replacing RFB modules with the decoder of Lawin Transformer.}
    \label{fig:mseg}
    \vspace{-0.5cm}
\end{figure}

  Fig.\ref{fig:mseg} depicts the original HarDNet-MSEG (a) and
  our enhanced model (b).
  Our enhancement includes modifying each HarDBlk module
  in the encoder backbone with a new HarDBlkV2 module and
  replacing RFB modules in the decoder with that of Lawin Transformer\cite{lawin}.
  
\subsection{HarDNetV2 -- Channel Balanced HarDNet}

\begin{figure}[!ht]
    \begin{minipage}[b]{.5\linewidth}
    \centering
    \subfloat[]{
    \includegraphics[width=8cm]{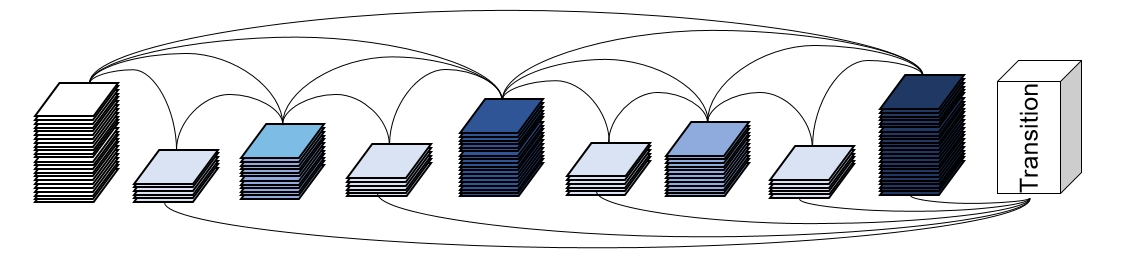}}
    
    \subfloat[]{ \includegraphics[width=8cm]{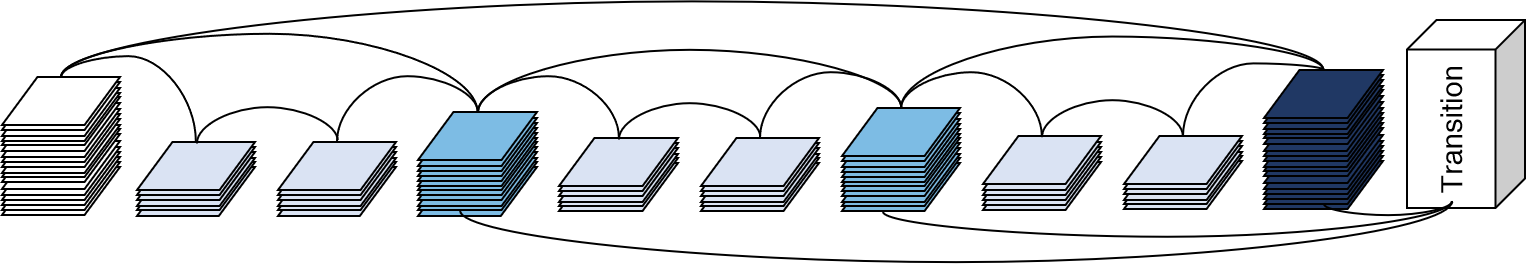}}
    \end{minipage}
    \medskip
    \begin{minipage}[b]{.7\linewidth}
    \centering
     \subfloat[]{ \includegraphics[width=3cm]{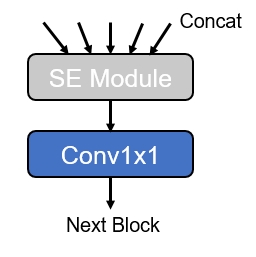}}
    \end{minipage}
    % \subfloat[]{ \includegraphics[width=2cm]{picture/Transition_Layer.png}}

    \caption{(a) HarDNet's basic building HarDBlock (n=8);
    (b) Enhanced basic building block HarDBlockV2 (n=9);
    (c) Transition Layer}
    \label{fig:block}
    %\vspace{-0.5cm}
\end{figure}

  % design principle
  HarDNet-MSEG's backbone consists of basic building blocks called HarDBlock.
  Our enhanced backbone incorporates ideas
  from CSPNet and ShuffleNetV2,
  and we call it HarDBlockV2 as depicted in Fig. \ref{fig:block}.
  To achieve the best MACs over CIO ratio (MoC),
  we perform channel splitting on
  the outputs of a convolutional layer \textit{l}
  according to its number of output connections.
  This makes the number of input channels equal to
  the number of output channels for each Conv3x3 layer.
  According to the design principle of HarDNet,
  the amount of DRAM access could be reduced.
  
%   In addition, we double the number of output channels
%   of a convolutional layer \textit{l} if
%   it is connected to the block output. 
%   The upper portion of \textit{l} will be sent
%   to the down-stream connected convolutional layers,
%   while the lower portion will be appended to
%   the block output.
%   Though it will make the input/output channel ratio of some
%   convolutional layers unbalanced, it
%   reduces the amount of redundant information in the block output.
  
  In addition, we propose a new link pattern
  that simplifies the network architecture design.
  We build the inter-layer connections according to the factors of
  the desired block depth \textit{n}.
  For example, when n=9, its factors are 1, 3, and 9,
  so we create shortcuts to 1st, 3rd, and 9th convolutional layers.
  By doing so, the depth of a basic building block in HarDNetV2 is
  no longer constrained to the power of 2.
  Instead of 4, 8, or 16 employed by HarDBlock, 
  we choose block depth\textit{ n}=3, 9, and 15
  to build HarDBlockV2,
  resulting in less data movement with the same number of convolutional layers.
  
  In the transition layer,
  we add an SE attention module \cite{SE} after
  the block output as shown in
  Fig. \ref{fig:block}(c).
  Because the block output concatenates
  some outputs from preceding layers,
  the attention module facilitates utilization of
  multi-scale information.
  
\subsection{Decoder}

HarDNet-MSEG was designed for real-time application of
colonoscopy examination.
Therefore, it trades accuracy for speed.
For an accuracy-oriented non-real-time
task such as foot ulcer segmentation,
we choose a more powerful decoder to obtain higher accuracy.
The authors of Lawin Transformer~\cite{lawin} proposed
an attention mechanism called Large Window Attention.
It utilizes MLP Decoder~\cite{segformer}, MLP-Mixer~\cite{mlpmixer},
and Spatial Pyramid Pooling(SPP)~\cite{spp} to capture multi-scale features.
Its abundant scale and attention can
represent the segmentation result more precisely
than the RFB decoder of HarDNet-MSEG does.

\subsection{Model Ensemble}

To increase the inference accuracy,
our ensemble strategy adopts
5-fold cross validation and Test Time Augmentation(TTA).
The dataset is randomly partitioned into
five folds of 400 images each.
For each cross-fold iteration,
four folds are used for training and the remaining one for validation.
After five iterations, we obtain five sub-models.
%% 說明TTA過程
%% method: horizontal, vertical, and horizontal-vertical flip
%TTA is a technique to boost the model accuracy by using data augmentation in the inference phase. 
During inference, we perform TTA on each sub-model.
That is, to generate an additional image via
image flipping,
feed both the test images and the additional image
to these sub-models, and
then take the average of their outputs as our prediction results.

\subsection{Loss Function}

Our loss function for DFUC2022 segmentation challenge
is given in Eq. \eqref{eq1},
  which calculates the loss between the ground truth $G$,
  the output $O$ of our model,
  the output $D_i$ of the deep supervision, and
  the output $B$ of the boundary.
  \begin{equation}
    L = l(G,O) + \sum_{i}l(G,D_i) + l_{BCE}(G_B,B)
    \label{eq1}
  \end{equation}
  where \begin{math}l(G, O) = l_{BCE}^{w}(G, O)+l_{IoU}^{w}(G, O)\end{math}, and \begin{math}l_{IoU}^{w}\end{math} and \begin{math}l_{BCE}^{w}\end{math} denote
  weighted IoU loss and weighted BCE loss, respectively.
  These two functions have the same definition as that of \cite{f3net}.
  $l_{BCE}(G_B,B)$ calculates the loss between
  the prediction boundary and the ground truth boundary.

\subsection{Post-Processing}

%To output the final segmentation masks, 
We pass the output through
the Tanh function and
normalize the result into the range [0, 1] and round to \{0, 1\} to represent a mask.
The last step is hole filling.
We first flood-fill the mask prediction from point (0, 0), then invert it as invertmask.
Finally, we OR the original mask and invertmask to get the hole-filled image as our final mask images.

\section{Experiments}

\subsection{Setting}

%% HarDNet-DFUS
For Diabetic Foot Ulcer Image Segmentation, we train the proposed models on
a single NVIDIA Tesla V100 GPU.
The batch size is 6
and learning rate is 1e-4 with cosine annealing schedule.
Training the model for 300 epochs takes about 15 hours.
To keep their original aspect ratio,
the training images are zero-padded into square and
then resized to 512 $\times$ 512.
We also employed multi-scaling.
The image would be randomly resized into multiples of 64
between 384(512$\times$0.75) and 640(512$\times$1.25).

Data augmentation includes random vertical flipping,
horizontal flipping, cropping, shifting, scaling, rotation, coarse dropout, brightness changing, contrast changing, and Gaussian noise introduction.

Our measurement metric is dice coefficient,
which is widely used in segmentation task. 

% We named the architecture in this section by
% the format "Model-Backbone-Decoder" as shown in Table \ref{tab:model}.
% For example, HarDNetV2-CSP69-Lawin means it uses HarDNetV2,
% the backbone is CSP69 and
% the decoder is a Lawin decoder.

%\begin{table}[!ht]
%\centering 
%%\vspace{-0.5cm}
%\caption{Five network architectures with different combinations of backbone and decoder \label{tab:model}}
%\begin{tabular}{c|c|c}
%\hline \hline
%% Problems: 上面method在講HarDNetV2的時候沒有提到CSP & 53/69 差異 
%% encoder? backbone? version?
%Model Name &  Backbone &  Decoder \\ \hline \hline
%{HarDNet-MSEG }        &  HarDNet           & RFB \\ \hline
%{HarDNetV2-53-RFB }    &  HarDNetV2-53      & RFB \\ \hline
%{\bfseries{HarDNetV2-53-Lawin (HarDNet-DFUS) }}  &  HarDNetV2-53  & Lawin \\ \hline 
%{HarDNetV2-CSP69-RFB } &  HarDNetV2-CSP69   & RFB \\ \hline
%{HarDNetV2-CSP69-Lawin } &  HarDNetV2-CSP69   & Lawin \\ \hline
                                  
%\end{tabular}
%%\vspace{-0.5cm}
%\end{table}

\subsection{Dataset}

For Diabetic Foot Ulcer Image Segmentation, the DFUC2022 dataset \cite{cassidy2020dfuc2020,goyal2018dfunet,goyal2018robust,goyal2020recognition,kendrick2022translating,yap2021analysis}
is provided by the organizer of MICCAI 2022 Diabetic Foot Ulcer Challenge.
It contains 2,000 640$\times$480 images with single-class ulcer segmentation labels.

For colonoscopy polyp segmentation,  we use Kvasir-SEG \cite{kvasir_seg}, CVC-ClinicDB \cite{cvc_clinicdb}, CVC-ColonDB \cite{cvc_colondb}, ETIS-Larib Polyp DB \cite{ETIS}, and EndoScene \cite{EndoScene}.

%% if 最後沒有交有pretrain facemask的版本: 此段刪除
% In addition, we used Face Mask dataset\cite{FaceMask} to be our pre-training data. We assume that our model can learn the information about the scene in daily life from Face Mask dataset, so that the background is less likely to be predicted as the foot ulcer, and our model can be more accurate. Face Mask dataset contains 222 images with single-class face mask labels. The size of images is either 1280$\times$1920 or 1920$\times$1280.
%數量100 images are 1280$\times$1920
%數量122 images are 1920$\times$1280

\subsection{Experiment Results}

\subsubsection{MICCAI DFUC 2022 Challenge}
%% 待修 
The following evaluation results are our submissions
during the validation phase of DFUC2022.
First, we study the representation power of the
new backbone by simply replacing
HarDNet-MSEG's original backbone, HarDNet, with the new one, HarDNetV2,
while keeping everything else unchanged.
Table \ref{tab:backbone} shows our new backbone
gained 1\% accuracy while keeping a similar speed.

%%%Different backbone in HarDNet-MSEG
\begin{table}
\centering
\caption{Effectiveness of new backbone \label{tab:backbone}}
\begin{tabular}{l|c|c}
\hline \hline
Model & Dice & FPS  \\ \hline \hline
{ HarDNet-MSEG }       & 0.6553 &           { 108 } \\ \hline
{ HarDNetV2-53-RFB } & { \bfseries 0.6651 } & 104 \\ \hline
\end{tabular}
\end{table}

We then study the effectiveness of model ensemble.
Table \ref{tab:5_Fold} shows the mean dice improvement
after using 5-fold cross validation.
Five-fold ensemble
gives the new network 0.8\% accuracy gain.

%%%HarDNetv2 w/ 5-Fold
\begin{table}
\centering
\caption{Effectiveness of 5-fold cross-validation and ensemble \label{tab:5_Fold}}
\begin{tabular}{l|c|c}
\hline \hline
                            Model & 5-Fold     & Dice  \\ \hline \hline
\multirow{2}{*}{HarDNetV2-53-RFB} &            & 0.6651  \\ \cline{2-3} 
                                  & \checkmark & { \bfseries 0.6730 } \\ \hline
\end{tabular}
\end{table}

In Table \ref{tab:architecture_compare},
we further compare different combinations of backbones and decoders.
Two versions of new backbones, HarDNetV2-53 and HarDNetV2-CSP69, and
two versions of decoders, RFB module and Lawin decoder, are investigated.
We designate the best architecture HarDNet-DFUS, i.e.,
the one with HarDNetV2 backbone (53 convolution layers) and Lawin decoder.

%%%53/69 + RFB/Lawin
\begin{table}
\centering 
\caption{Results of different combinations of new backbone sizes and decoder types \label{tab:architecture_compare}}
\begin{tabular}{l|c|c}

\hline \hline
Model                   & 5-Fold     & Dice  \\ \hline \hline
HarDNetV2-53-RFB        & \checkmark & 0.6730   \\ \hline
HarDNetV2-CSP69-RFB     & \checkmark & 0.6842  \\ \hline
HarDNetV2-53-Lawin (HarDNet-DFUS)    & \checkmark & { \bfseries 0.6950 }   \\ \hline
HarDNetV2-CSP69-Lawin   & \checkmark & 0.6870  \\ \hline
\end{tabular}
\end{table}

\begin{figure}
    \centering 
    \includegraphics[width=8cm]{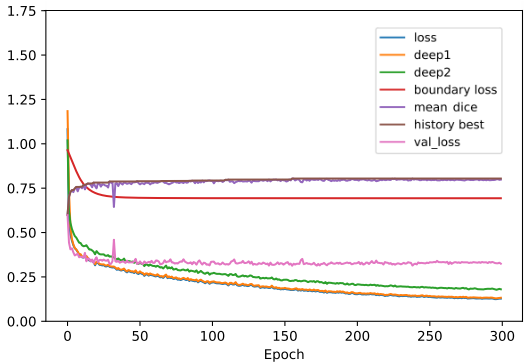}\label{fig:sub_ref_5} 

    \caption{Training process of HarDNet-DFUS (HarDNetV2-53-Lawin) in one of five folds.}
    \label{fig:train}
    \vspace{-0.5cm}
\end{figure}

Fig. \ref{fig:train} shows the loss and dice of HarDNet-DFUS
during training in one of five folds.
We plot the loss, deep supervision loss(deep1 and deep2),
boundary loss(boundary loss), mean dice(dice),
the current best dice(best dice) and the validation loss(val\_loss) at each epoch.

%%%

%The result of different deep-supervision shown in Table \ref{tab:deep}, we also experiment an additional boundary loss.
As shown in Table \ref{tab:deep}
we experiment with different deep supervision.
There are two deep supervision losses and one boundary loss.
We can see deep supervision loss works
when we take more than one to join the training.

\begin{table}
\centering
\caption{Segmentation accuracy of HarDNet-DFUS using different combinations of loss functions\label{tab:deep}}
\begin{tabular}{ccc|c}
\hline \hline
{ Deep1 }      & { Deep2 }      & { Boundary }   & { Dice }  \\ \hline \hline
{  }           &                &                & 0.6915      \\ \hline
{ \checkmark } &                &                & 0.6852      \\ \hline
{ \checkmark } & { \checkmark } &                & 0.6950      \\ \hline
{ \checkmark } &                & { \checkmark } & {\bfseries 0.7001}      \\ \hline
{ \checkmark } & { \checkmark } & { \checkmark } & 0.6927      \\ \hline
\end{tabular}
\end{table}

Table \ref{tab:tta} compares the effect of test time augmentation (TTA) on
different combinations of deep supervision.
TTA includes none, horizontal flip, vertical flip,
and horizontal flip with a vertical flip.
It increases the accuracy in some cases.
However, its effect is not robust.

%%%TTA比較
\begin{table}[!ht] 
\centering
\caption{Effect of Different Test Time Augmentations in HarDNet-DFUS\label{tab:tta}}
\begin{tabular}{l|c|c}
\hline \hline
                        Model   & TTA Method            & Dice  \\ \hline \hline
\multirow{4}{*}{HarDNet-DFUS} & none                    & 0.6920  \\ \cline{2-3}
                                & horizontal            & 0.6947 \\ \cline{2-3}
                                & vertical              & 0.6931 \\ \cline{2-3}
                                & horizontal+vertical   & { \bfseries 0.6975} \\ \hline

\multirow{4}{*}{HarDNet-DFUS+Deep1+Deep2} &        none & 0.6950  \\ \cline{2-3}
                                & horizontal            & 0.6958 \\ \cline{2-3}
                                & vertical              & { \bfseries 0.6992} \\ \cline{2-3}
                                & horizontal+vertical   & 0.6943 \\ \hline
                                
\multirow{4}{*}{HarDNet-DFUS+Deep1+Boundary} &    none  & { \bfseries 0.7001}  \\ \cline{2-3}
                                & horizontal            & 0.6981 \\ \cline{2-3}
                                & vertical              & 0.6934 \\ \cline{2-3}
                                & horizontal+vertical   & 0.6928 \\ \hline

\multirow{4}{*}{HarDNet-DFUS+Deep1+Deep2+Boundary} & none& 0.6927  \\ \cline{2-3}
                                & horizontal            & 0.6994 \\ \cline{2-3}
                                & vertical              & { \bfseries 0.7063} \\ \cline{2-3}
                                & horizontal+vertical   & 0.6985 \\ \hline
\end{tabular}
\end{table}

We observe some small values being classified as positive after being compressed by the Sigmoid function, but not by the Tanh function.
So we compare Sigmoid and Tanh and show the results in Table \ref{tab:tanh}. Generally, Tanh gives us better results.

%%%tanh v.s. sigmoid
\begin{table}[!ht] 
\centering
\caption{Effects of prediction compression(Sigmoid vs Tanh) in HarDNet-DFUS\label{tab:tanh}}
\begin{tabular}{l|c|c}
\hline \hline
                        Model   & \makecell[c]{Compressing \\ Method}   & Dice  \\ \hline \hline
\multirow{2}{*}{HarDNet-DFUS+Deep1+Boundary}   & Sigmoid       & 0.6752  \\ \cline{2-3}
                                & Tanh          & { \bfseries 0.7001} \\ \hline

\multirow{2}{*}{HarDNet-DFUS+Deep1+Boundary (w/ hflip)}   & Sigmoid       & 0.6834  \\ \cline{2-3}
                                & Tanh          & { \bfseries 0.6981} \\ \hline
                                
\multirow{2}{*}{HarDNet-DFUS+Deep1+Deep2+Boundary (w/ hflip)}   & Sigmoid       & 0.6950  \\ \cline{2-3}
                                & Tanh          & { \bfseries 0.6994} \\ \hline

\multirow{2}{*}{HarDNet-DFUS+Deep1+Deep2+Boundary (w/ vflip)}   & Sigmoid       & 0.7029  \\ \cline{2-3}
                                & Tanh          & { \bfseries 0.7063} \\ \hline
                                
\multirow{2}{*}{HarDNet-DFUS+Deep1+Deep2+Boundary (w/ vflip \& hflip)}   & Sigmoid       & { \bfseries 0.6995}  \\ \cline{2-3}
                                & Tanh          & 0.6985 \\ \hline
\end{tabular}
\end{table}

For the validation phase of 2022 MICCAI DFU Challenge,
HarDNet-DFUS achieves 0.7063 mean dice and ranked third among 21 participating teams. 
Table \ref{tab:testing} shows the results of our five submissions
during the final testing phase.
HarDNet-DFUS achieves 0.7287 mean dice and ranked first among all teams.
Rather than the poly-oriented HarDNet-MSEG, HarDNet-DFUS has 5\% higher mean dice for the DFUC task. 
%we have also tried using the polyp-oriented HarDNet-MSEG for the DFUC task.
%The mean dice is 5\% lower.

%%%testing phase results
\begin{table}[!ht]
\centering
\caption{Results of five Submissions of HarDNet-DFUS in the Final Testing Phase of 2022 MICCAI DFUC\label{tab:testing}}
\begin{tabular}{l|c}
\hline \hline
{ Model } & { Dice }  \\ \hline \hline
{ HarDNet-DFUS+Deep1+Boundary }                & 0.7237  \\ \hline
{ HarDNet-DFUS+Deep1+Boundary (w/ hflip) }     & 0.7243  \\ \hline
{ HarDNet-DFUS+Deep1+Deep2+Boundary (w/ hflip) }  & 0.7273  \\ \hline
{ HarDNet-DFUS+Deep1+Deep2+Boundary (w/ vflip) }  & 0.7275  \\ \hline
{ HarDNet-DFUS+Deep1+Deep2+Boundary (w/ vhflip) } & { \bfseries 0.7287}  \\ \hline
\end{tabular}
\end{table}

\subsubsection{HarDNet-DFUS for Polyp Segmentation}

We study how HarDNet-DFUS perform on the task of polyp segmentation following the training and experiment setup of HarDNet-MSEG\cite{HarDNet-MSEG}.

In HarDNet-MSEG\cite{HarDNet-MSEG}, 1450 training images were used,
including 900 images in Kvasir-SEG and 550 images in CVC-ClinicDB.
And the testing set has 5 datasets including Kvasir-SEG, CVC-ClinicDB, CVC-ColonDB, ETIS-Larib Polyp DB, and EndoScene.
CVC-T is the testing data for EndoScene.
Our training input size is 384x384.
We train HarDNet-DFUS with AdamW optimizer for 300 epochs and the learning rate is set to 1e-4.
The quantitative results of the five popular datasets are shown in Table \ref{tab:polyp}.
The results show that HarDNet-DFUS delivers better performance than HarDNet-MSEG on four datasets and retains real-time performance
despite using a more complex decoder. 
%%%Quantitative results on Kvasir, CVC-ClinicDB, CVC-ColonDB, ETIS, and CVC-T dataset, comparing with HarDNet-MSEG.
\begin{table}
\centering
\caption{HarDNet-DFUS(+Deep1+Deep2+Boundary) improved over HarDNet-MSEG on Popular Polyp Segmentation Datasets. \label{tab:polyp}}
\begin{tabular}{l|c|c|c|c|c|c|c}
\hline \hline
Model              & Kvasir     & ClinicDB             & ColonDB              & ETIS                 & CVC-T   &FPS   \\ \hline \hline
{ HarDNet-MSEG }   & 0.912      & 0.932                & 0.731                & 0.677                & { \bfseries 0.887} & { \bfseries 108} \\ \hline
{ HarDNet-DFUS}    & { \bfseries 0.918}      & { \bfseries 0.939}   & { \bfseries 0.774}   & { \bfseries 0.730}   & 0.876   & 30 \\ \hline
                                  
\end{tabular}
\end{table}

\section{Conclusion and Future Work}

For the task of diabetic foot ulcer segmentation,
we have proposed enhancing the previously state-of-the-art HarDNet-MSEG
polyp segmentation network
with a new backbone and a more powerful decoder.
We call the new network HarDNet-DFUS.
Five-fold cross validation, deep supervision, boundary supervision, and
test time augmentation together contribute to about 5\% improvement
in mean dice compared with the original HarDNet-MSEG.
We have participated in the 2022 MICCAI DFUC Challenge and 
have been awarded the first place winner.
HarDNet-DFUS also deliver excellent performance for
colonoscopy polyp segmentation.
Compared with HarDNet-MSEG, it has better accuracy on
Kvasir, CVC-ClinicDB, CVC-ColonDB, and ETIS datasets
while retaining real-time speed.

In the future, we would like to deploy HarDNet-DFUS in
clinical fields and expand its application scope
to more medical imaging tasks.

\subsubsection{Acknowledgements}

This research is partially supported by
the Ministry of Science and Technology (MOST) of Taiwan.
We thank the National Center for High-performance Computing (NCHC) for
computational and storage resources.
We would also like to thank Professor Tzu-Chen Dorothy Yen and
Professor Chang-Fu Kuo of Chang-Gang Memorial Hospital for their advice.

\bibliographystyle{splncs04}
\bibliography{ref}

\end{document}